\documentclass{ws-mpla}

\begin{document}

\markboth{M.~Risse \& P.~Homola}
{Search for UHE photons using air showers}

\catchline{}{}{}{}{}

\title{SEARCH FOR ULTRA-HIGH ENERGY PHOTONS \\ USING AIR SHOWERS}

\author{\footnotesize MARKUS RISSE}

\address{University of Wuppertal,
Department of Physics \\
Gau\ss str.~20, 
42097 Wuppertal, Germany \\
risse@physik.uni-wuppertal.de}

\author{PIOTR HOMOLA}

\address{H.~Niewodnicza\'nski Institute of Nuclear Physics PAN \\
ul.~Radzikowskiego 152, 31-342 Krak\'ow, Poland \\
piotr.homola@ifj.edu.pl
}

\maketitle

\pub{Received (Day Month Year)}{Revised (Day Month Year)}

\begin{abstract}
The observation of photons with energies above $10^{18}$~eV 
would open a new window in cosmic-ray research, with possible
impact on astrophysics, particle physics, cosmology and
fundamental physics.
Current and planned air shower experiments, particularly the
Pierre Auger Observatory, offer an unprecedented
opportunity to search for such photons and to complement efforts
of multi-messenger observations of the universe.
We summarize motivation, achievements,
and prospects of the search for ultra-high energy photons.

\end{abstract}

\ccode{96.40.Pq,96.40.-z,13.85.-t,13.85.Tp}

\section{Introduction}

Photons are the main messenger particles for exploring the universe.
Over the last decades, the wavelength range of photon observation
was dramatically expanded and stretches now from radiowaves to
high-energy gamma rays. 
The current maximum energy of photons observed is 
$\sim$$10^{14}$~eV.\cite{weekes}

In this short review, we focus on the search for photons of
energies above $10^{18}$~eV up to the highest energy
(few times $10^{20}$~eV) by measuring particle cascades
(air showers) initiated in the atmosphere of the Earth.
An observation of such ultra-high energy (UHE) photons, several
orders of magnitude in excess of currently observed photon energies,
would open a new window in cosmic-ray research with significant
impacts on related research fields.
Giant air shower experiments, most prominently 
the Pierre Auger Observatory\cite{auger}, are unique tools to explore
this photon energy range.
The expected sensitivity of the Auger Observatory after installation
of a large northern array reaches the photon fluxes predicted in
more conservative scenarios of cosmic-ray origin, making an
observation of UHE photons feasible.

UHE photons are also inherently linked to other messenger particles
such as charged cosmic 
rays\cite{reviews_cr1,reviews_cr2,reviews_cr3,reviews_cr4}
and neutrinos\cite{reviews_nu1,reviews_nu2}.
The search for UHE photons complements current experimental efforts
towards multi-messenger observations of the universe.

The structure of the paper follows the possible ``life'' of
UHE photons. Production scenarios and propagation 
of photons as well as their predicted fluxes at Earth
are shortly described in Section~\ref{sec-cr}.
Geomagnetic and atmospheric cascading and
the challenge of identifying photon showers are discussed in
Section~\ref{sec-eas}.
Current observational results and future prospects
are summarized in Section~\ref{sec-exp}.
Throughout Sections~\ref{sec-cr}$-$\ref{sec-exp},
conceptual issues important for the data interpretation are
pointed out.
The possible impact of photon searches and detections on different
research fields is briefly outlined in Section~\ref{sec-impact}.
Conclusions are given in Section~\ref{sec-summ}.

\section{UHE Photons As Cosmic Rays}
\label{sec-cr}

{\bf Production.}
Though at different levels, most models predict UHE photons,
mainly from the decay of neutral pions produced previously by
a ``primary process'',
\begin{equation}
\mbox{primary process} ~\rightarrow
 ~\pi^0 ~(+ \pi^\pm) + ... 
~\rightarrow~ \gamma_{_{\rm UHE}} ~(+ \nu_{_{\rm UHE}}) + ...
\label{eq-production}
\end{equation}
In non-acceleration models\cite{bhat-sigl}, the primary process is given
by the decay or annihilation of primordial relics such as topological
defects\cite{td1,td2} (TD) or super heavy dark 
matter\cite{shdm1,shdm2,shdm3,shdm4} (SHDM).
From considerations of QCD fragmentation,\cite{frag1,frag2,frag3,berez04}
copious photons
are then expected to be produced.\cite{sarkar03,models,ellis}
In the Z-burst scenario\cite{zb1,zb2,zb3} (ZB), photons are generated via the resonant
production of Z bosons by UHE 
neutrinos annihilating on the relic neutrino background.

\begin{figure}[t]
\centering
\begin{minipage}[c]{.45\textwidth}
\centering
\leftline{\psfig{file=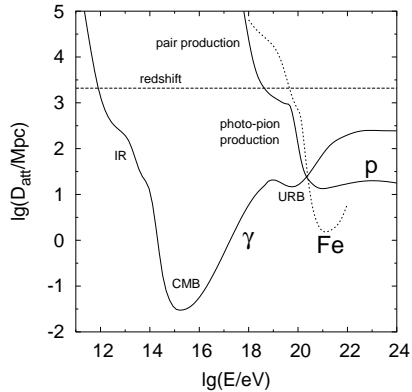,width=3.0in}}
\end{minipage}
\begin{minipage}[c]{.45\textwidth}
\centering
\caption{Energy loss length of photons for interactions with
infrared (IR), cosmic microwave (CMB) and universal radio (URB)
backgrounds. Uncertainties exist for the URB (see text) and IR background.
For comparison, curves for protons and iron nuclei are added.
Values are compiled from Refs.\protect\cite{bhat-sigl,losslengths1,losslengths2}.
\protect\label{fig-mfp}}
\end{minipage}
\end{figure}

In more ``conventional'' cosmic-ray models, nuclear primaries are
accelerated at suitable astrophysical sites to ultra-high energy.
UHE photons can be produced during propagation by the GZK-type 
process\cite{gzk1,gzk2}
of resonant photo-pion production of UHE nucleons with the cosmic microwave
background.
The energies of these ``GZK photons'' are typically a factor $\sim$10 below the
primary nucleon energy.
An enhanced production of photons $>$$10^{18}$~eV can occur from
nuclear primaries passing near the galactic center region.\cite{stecker06}
%
\\ \\
{\bf Propagation.}
UHE photons can initiate electromagnetic
cascades by interacting with background radiation fields,
\begin{equation}
\gamma_{_{\rm UHE}} + \gamma_{_{\rm background}} ~\rightarrow
 ~e^\pm ~\rightarrow  ... ~\rightarrow \gamma_{_{\rm GeV-TeV}} + ...
\label{eq-propagation}
\end{equation}
Significant uncertainties exist for the low-frequency (few MHz) radio
background and (for $e^\pm$ propagation)
extragalactic magnetic fields.\cite{sarkar03,models}
Typical energy loss lengths assumed for UHE photons range between
7--15~Mpc at $10^{19}$~eV and 5--30~Mpc at $10^{20}$~eV.
In SHDM models, the relic particles are clustered as cold dark
matter in our Galaxy, and UHE photons, as their decay products, would
be observed at Earth with little processing.
In TD and ZB models, UHE photons are injected at larger distance
from the Earth, and part of these photons can cascade to lower energy.
The electromagnetic cascade stops at GeV--TeV energies where the universe
becomes increasingly transparent for photons, see Figure~\ref{fig-mfp}.
%
\\ \\
{\bf Flux predictions.}
In addition to the mentioned
uncertainties of UHE photon propagation, there are free
parameters within the theoretical source models such as
density and lifetime of primordial relics in non-acceleration models,
injection spectrum in acceleration models, source
distribution, etc.
To calculate the (range of) photon fluxes predicted by a model,
the measured cosmic-ray spectrum is usually taken as a
constraint.
The question arises which spectrum
to use. Firstly, the absolute UHE cosmic-ray 
flux differs between the experiments by a factor $\sim$2.
This problem may be reduced by regarding the fraction
of photons in the cosmic-ray flux rather than absolute rates.
Secondly, it is unclear so far whether
a suppression of the flux (``GZK cutoff''\cite{gzk1,gzk2})
above $E_{\rm GZK} \sim 6 \times 10^{19}$~eV
exists (as indicated by HiRes data\cite{hires-gzk}) 
or not (as indicated by AGASA data\cite{agasa-gzk}).
The shape of the assumed energy spectrum can affect also
the predicted fraction of photons.\cite{models,aloisio}

\begin{figure}[t]
\centerline{\psfig{file=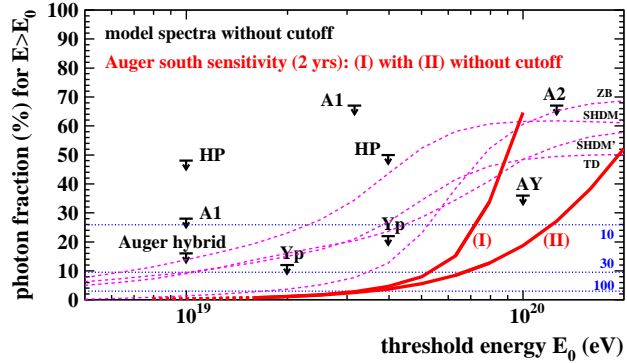,width=3.3in}}
\vspace*{8pt}
\caption{
Fraction of photons in the integral cosmic-ray flux as
a function of the threshold energy.
The predictions of non-acceleration models (dashed lines:
ZB, SHDM, TD from~\protect\cite{models}, 
SHDM' from~\protect\cite{ellis})
follow a spectrum without flux suppression above $E_{\rm GZK}$.
Experimental limits (Section~\ref{sec-status};
all limits at 95\% c.l.) are reported from
Auger hybrid observations (Auger hybrid)\protect\cite{augerphoton},
Haverah Park measurements (HP)\protect\cite{ave1,ave2}, AGASA data
(A1)\protect\cite{shinozaki}, (A2)\protect\cite{risse05},
Yakutsk data (Yp, preliminary)\protect\cite{yakutsk}, and
a combination of AGASA and Yakutsk data
(AY)\protect\cite{troitsky1,troitsky2}.
To illustrate the minimum number of events required to place
a (95\% c.l.) limit to a certain photon fraction, horizontal dotted
lines are shown with the number of events assigned.
An estimate (Section~\ref{sec-prosp}) of the sensitivity of the southern
Auger Observatory for two years
of operation is given for the two cases that the real spectrum
(I) does have, (II) does not have a cutoff.
The data from the southern Auger Observatory allow a stringent test of
non-acceleration scenarios even if the real spectrum had a cutoff.
\protect\label{fig-uplim}}
\end{figure}

Non-acceleration models are usually calculated assuming
a spectrum without flux suppression.
These models aim at explaining the highest-energy end of
the spectrum only.\cite{aloisio}
For energies up to $\sim$$E_{\rm GZK}$, a ``conventional''
component of nuclear primaries must be assumed.
Then, a common prediction of the models is a dominant
photon component at $\sim$$10^{20}$~eV, see Figure~\ref{fig-uplim}.
A reduction of the predicted photon fraction may formally
be achieved by introducing a flux suppression above $E_{\rm GZK}$
which increases the relative importance of the ``conventional''
nuclear component. However, at the same time this reduces the
relevance of non-acceleration models to explain any 
observed events at all.

For acceleration models, in turn, it seems more natural to assume
a spectrum with flux suppression
as no nearby astrophysical sources could be identified by 
now.
The predicted photon fluxes are typically relatively small.
Fractions of order $\sim$0.1\% were obtained in scenarios assuming
nucleon sources,\cite{models,sigl06} with a considerable
range of uncertainty (see Figure~\ref{fig-uplim}).
Larger photon fractions may emerge for specific
assumptions on source features, particularly when trying to
reproduce a spectrum without flux suppression.\cite{models}
In case of primary nuclei, the rate of GZK produced photons may be
suppressed due to the higher total energy of the nucleus required.
Further investigations of such scenarios and of the relation between
the photon flux and model parameters are desirable.

A caveat seems in place. The reconstruction of the energy spectrum
from shower observations itself requires, to some extent, the
fraction of photons as an input (Section~\ref{sub-det}).
Usually, no contribution from primary photons is assumed in the
reconstruction.
For model predictions, a self-consistent comparison to data
is required. 

\section{UHE Photons as Air Shower Primaries}
\label{sec-eas}

At the Earth, photons initiate almost purely
electromagnetic showers via pair production and bremsstrahlung.
Additional processes are important at highest energy.

\subsection{Specific high-energy processes}
\label{sub-process}

\begin{figure}[t]
\centerline{
\psfig{file=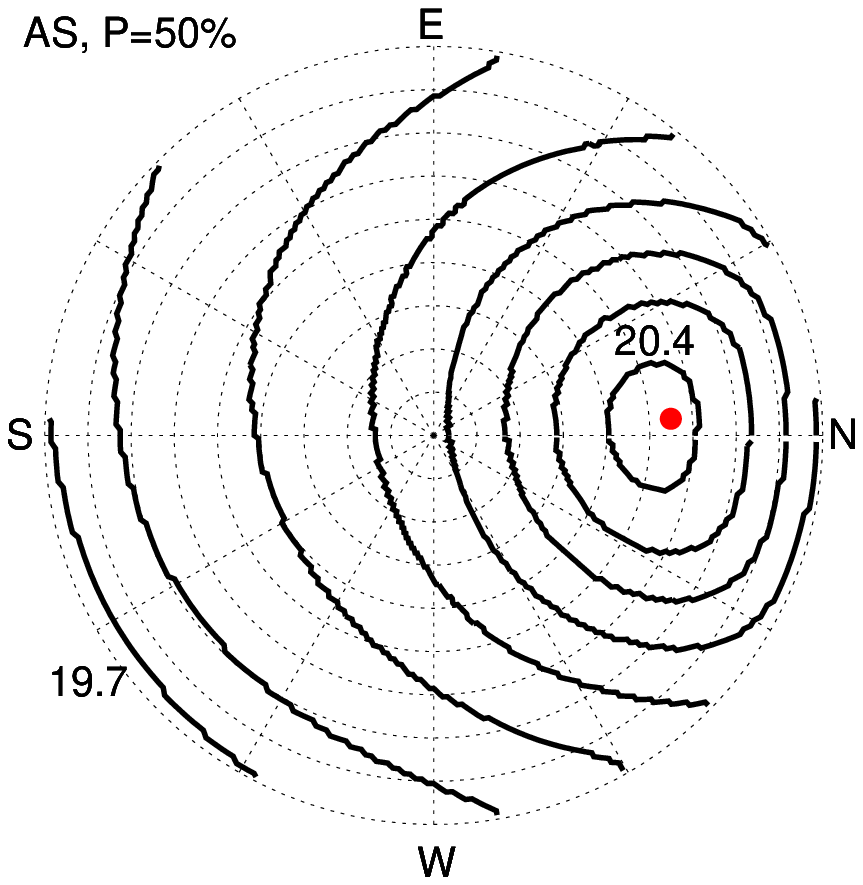,width=2.3in}
\psfig{file=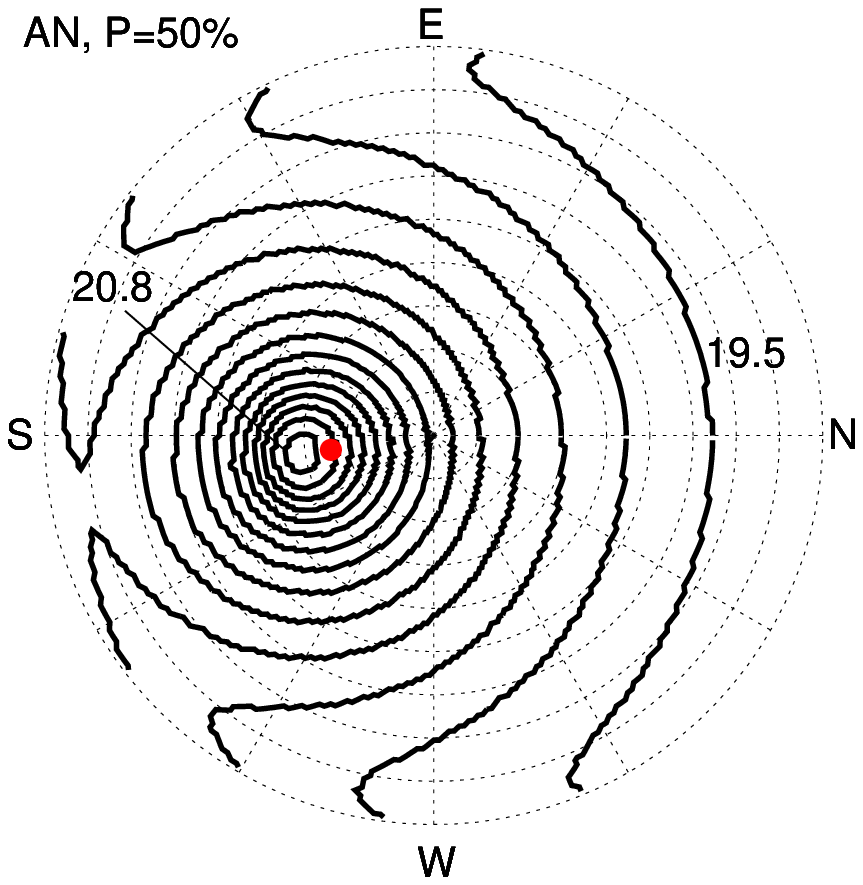,width=2.3in}}
\vspace*{8pt}
\caption{
Sky maps of photon energies at which the conversion probability
is $P_{\rm conv} = 50\%$ for the southern (left) and planned northern 
(right) Auger sites.\protect\cite{ns_photons}
Contour lines are given with a stepsize of
$\Delta \lg (E/$eV) = 0.1, minimum and maximum
energies are assigned.
Azimuthal directions are labeled (``E'' for East etc.).
Zenith angles are given as concentric
circles in 10$^\circ$ steps (zenith in the center).
With increasing energy, the sky fraction grows where the preshower
process is important. At Auger North, preshowering starts
at factor $\sim$2 smaller energy compared to Auger South
due to the stronger local magnetic field. Close to the pointing direction of
the local magnetic field (indicated by a red dot),
the highest energies are required for conversion to occur. Accordingly,
the sky patterns are shifted in local coordinates between
the sites. Despite the larger magnetic field at Auger North,
higher energies are needed to reach $P_{\rm conv} = 50\%$ close
to the local field direction. This is connected to the field lines
being less curved with altitude at Auger North. The differences between
the site characteristics can be exploited in photon
searches.\protect\cite{ns_photons}
\protect\label{fig-conv}}
\end{figure}

{\bf Preshower effect.}
Contrary to nuclear primaries, $\sim$10$^{20}$~eV photons can convert
in the geomagnetic field to an $e^\pm$ pair which then emits synchrotron
photons.\cite{presh1,presh2,presh3,presh4,presh5,presh6,daugherty83}
Instead of a single UHE photon, a bunch of lower-energy electromagnetic
particles, called ``preshower'', enters the atmosphere
with important consequences for the shower development.
The local differential probability of photon conversion 
as well as the probability distribution of
synchrotron photons emitted by the electrons, depend on the parameter
\begin{equation}
\chi=\frac{E}{mc^2}\frac{B_{\perp}}{B_{\rm c}}~, 
~~B_{\rm c}\sim 4.414 \times 10^{13}~{\rm G}
\end{equation}
where $E$ is the energy of the parent particle (photon or electron),
$m$ the electron mass, $B_{\rm c}$ a constant,
and $B_{\perp}$ is the local magnetic field component transverse to the
direction of the particle's motion.
Since $B_{\perp}$ is involved, preshower characteristics depend strongly
on the specific trajectory through the magnetosphere and, thus, on the
arrival direction and the experimental site (see Figure~\ref{fig-conv}).
The probability $P_{\rm conv}$ of a photon to convert
in the Earth's magnetic field results from an integration along
the particle trajectory; for $P_{\rm conv} \rightarrow 1$,
photons would almost surely undergo geomagnetic cascading.
Non-negligible probabilities ($P_{\rm conv} \sim$ 10\%) are
usually obtained if values $\chi > 0.5$ are reached along the
trajectory, corresponding to photon energies above 
 2--4 $\times$ $10^{19}$~eV depending on the site.

The energy partition in the produced $e^\pm$ pair is usually symmetric
(ratio of energies $<$2), with the probability of asymmetric
share increasing with $\chi$.\cite{daugherty83,coraddu02}
The spectrum of synchrotron photons 
becomes harder for larger values of $\chi$.
Secondary photons of sufficiently high
energy can convert again.

A typical preshower from a 10$^{20}$~eV photon 
starts at $\sim$1000~km altitude and enters the
atmosphere ($\sim$100~km altitude) 
with one or a few $e^\pm$ pairs around 10$^{18}$~eV
and a large number ($\sim$500) of photons.
The photon energies extend over several decades, with a
few photons around 10$^{19}$~eV.
Significant fluctuations around these averages may occur,
particularly when the conversion takes place at low altitudes.
The spread of preshower particles in transverse distance and arrival time
is well below current detector resolutions; the subsequent air shower
is observed as one event.\cite{cpc}

To simulate the preshower effect, Monte Carlo codes have been developed.
The PRESHOWER\cite{cpc} program is available as a standalone tool
and is also linked to the CORSIKA\cite{heck} and
CONEX\cite{conex} shower codes. An independently developed
preshower program (MaGICS\cite{magics}) is available within 
the shower code AIRES\cite{aires}.
A realistic model of the geomagnetic field such as the
IGRF\cite{igrf} model,
is an essential ingredient of precise calculations.
As an example, a simple dipole model fails to reproduce the factor $\sim$2
difference in local field strength at the southern and northern
Auger sites.
\\ \\
{\bf LPM effect.}
In a medium,
the Bethe-Heitler cross-section\cite{bh} for pair production by photons
($\sigma_{\rm BH} \approx 0.51$ b in air\cite{pdg06}) can be reduced
due to destructive interference from several scattering centers.
This so-called LPM effect (Landau and 
Pomeranchuk\cite{lanpom53a,lanpom53b}, Migdal\cite{mig56})
is confirmed by experiments, see Refs.\cite{klein1,klein2} for reviews.
With
\begin{equation}
\label{eq-kappa}
\kappa = \frac{E_{\gamma}E_{LPM}}{E_e(E_{\gamma}-E_e)},
~~E_{\rm LPM}=\frac{m^2c^3\alpha X_0}{4\pi\hbar\rho}\approx (7.7 \textrm{~TeV/cm})
 \times \frac{X_0}{\rho},
\end{equation}
the reduced cross-section $\sigma_{LPM}$ can for $\kappa$$<$1 be
approximated\cite{klein1,klein2} by
$\sigma_{LPM} = \sigma_{BH}\sqrt{\kappa} \propto (\rho E_{\gamma})^{-\frac{1}{2}}$,
with photon energy $E_{\gamma}$, electron energy $E_e$,
radiation length $X_0 \sim 37$~g cm$^{-2}$ in air,
density $\rho$ of the medium, and electron mass $m$.
It follows from Eq.~(\ref{eq-kappa}) that
the reduction is largest for conversion to a symmetric electron pair
($E_e \approx$ $E_{\gamma}/2$). The cross-section for producing a
highly asymmetric pair ($E_e / E_{\gamma} \rightarrow $ 0 or 1)
changes only slightly.
In a similar way, also bremsstrahlung is suppressed.\cite{pdg06,hansen}
Numerical examples (see e.g. Ref.\cite{cillis}) are
$E_{\rm LPM}\sim 2.8 \times 10^{17}$~eV at 300 m a.s.l.~and
$\sim 10^{19}$~eV
in the upper atmosphere.

The LPM effect delays the development of an air shower initiated by
a single UHE photon as those processes are suppressed that
degrade the energy carried by individual high-energy particles.
Fluctuations can be very
large due to a positive correlation of the reduction of $\sigma_{LPM}$
since $\sigma_{LPM}(X_2) < \sigma_{LPM}(X_1)$ for depths $X_2$$>$$X_1$.
The LPM effect is accounted for in 
AIRES, CORSIKA, and CONEX.
\\ \\
{\bf Photonuclear interactions.}
Photon-initiated cascades are almost purely electromagnetic ones.
Production of muon pairs is suppressed by $(m_e / m_\mu)^2$.
The cross-section for photonuclear interactions which mainly transfer
energy to secondary hadrons (and these subsequently to muons), is expected
to be $\sim$10~mb at $10^{19}$~eV and thus more than two orders of magnitude
below $\sigma_{\rm BH}$ (which, however, can be reduced by the LPM effect).

One important consequence is that our lack of knowledge of hadron dynamics at
high-energy $-$ a major limitation for conclusions about the nuclear
composition~$-$ is of much smaller impact when calculating photon showers.
For instance, depths of shower maxima differ by 30$-$40 g~cm$^{-2}$
between SIBYLL\cite{sib21} and QGSJET\cite{qgs01,qgs2}
for UHE protons (see Section~\ref{sub-feat}, Figure~\ref{fig-xmaxvse})
but by less than $\sim$~5~g~cm$^{-2}$ for UHE photons\cite{augerphoton}.
For photon searches, data can be compared to photon simulations only,
without subtracting a background from nuclear primaries.

As another consequence, certain photon shower observables, in particular
the number of secondary muons, are
sensitive to UHE extrapolations of the photonuclear cross-section.
Different extrapolations of the photon-proton cross-section exist
in the literature, see e.g. Refs.\cite{rissec2cr,strikman}
for compilations.
For UHE photon searches, it is important to know whether
the cross-section could grossly exceed an extrapolation such as provided by
the Particle Data Group (PDG),\cite{pdg1,pdg2} since larger values
mean photon showers were actually more similar to those of nuclear primaries;
an absence of UHE photons may then erroneously be concluded from data.
Using a dipole formalism and taking unitarity constraints into account,
values of the UHE cross-section that exceed the PDG extrapolation
by more than $\sim$80\% (less at smaller energy) were found to be
disfavored.\cite{strikman}
It should be noted that for shower simulations, finally the 
photon-{\it air} cross-section is needed, which requires an account for
nuclear effects (see e.g.~Ref.\cite{strikman}).
From simulations with modified cross-sections,
uncertainties for photon showers
of $\sim$10~g~cm$^{-2}$ in $X_{\rm max}$ (depth of shower maximum)
and $\sim$15\% in $N_\mu$ 
(number of muons at ground) were estimated.\cite{rissec2cr}
Assuming an extrapolation such as from Ref.\cite{donland} which exceeds
the PDG fit by a factor $\sim$10 at $10^{20}$~eV,
a reduction of $X_{\rm max}$ by $\sim$100~g~cm$^{-2}$ is possible
for unconverted photons.
Already with a small sample of observed photon showers, such a scenario
can be distinguished from more ``standard'' ones.

\subsection{Features of photon-induced showers}
\label{sub-feat}

\begin{figure}[t]
\centering
\begin{minipage}[c]{.6\textwidth}
\centering
\centerline{\psfig{file=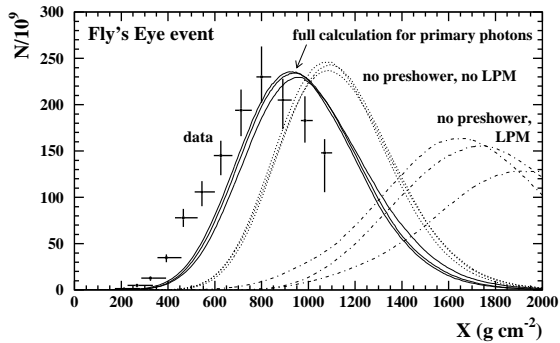,width=3.2in}}
\end{minipage}
\begin{minipage}[c]{.35\textwidth}
\centering
\caption{
Observed profile of the $3.2 \times 10^{20}$~eV Fly's Eye
event\protect\cite{bird} (data points) compared to
simulations \protect\cite{fe04} assuming primary photons
when neglecting both preshower and LPM effect (dotted line),
accounting for LPM effect only (dotted-dashed line),
and accounting for both effects (solid line).
Data points are correlated in $X$. Photons differ by
$\sim$1.5$\sigma$ from the data which corresponds to
a $\sim$13\% chance probability.\protect\cite{fe04,halzen_fe}
\protect\label{fig-profiles}
}
\end{minipage}
\end{figure}

Shower profiles calculated with the preshower and LPM effects
switched on/off are shown in Figure~\ref{fig-profiles}
for the conditions of the $3.2 \times 10^{20}$ Fly's Eye event\cite{bird}.
Both effects compete with each other:
the LPM effect increases the average $X_{\rm max}$ and its event-by-event
fluctuations. The preshower effect, in turn,
reduces both quantities.

\begin{figure}[t]
\begin{minipage}[l]{.57\textwidth}
\centerline{\psfig{file=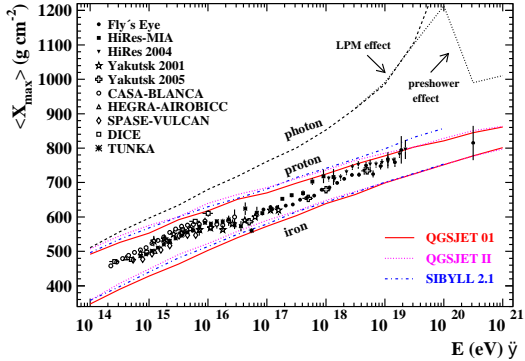,width=2.7in}}
\end{minipage}
\begin{minipage}[c]{.42\textwidth}
\centering
\caption{
Average depth of shower maximum $<$$X_{\rm max}$$>$ versus energy
simulated for primary photons, protons and iron nuclei.
The impact of the LPM and preshower effects on $<$$X_{\rm max}$$>$
is visible.
The splitting of the photon line at $\sim$3$\times 10^{19}$~eV
indicates that $<$$X_{\rm max}^\gamma$$>$ above these energies
depends also on the specific trajectory through the geo\-magnetic field.
For nuclear
primaries, calculations for different hadronic interaction models
are displayed (SIBYLL 2.1\protect\cite{sib21},
QGSJET 01,\protect\cite{qgs01}
QGSJET II \protect\cite{qgs2}).
See Ref.\protect\cite{xmax-heck} for references to the
experimental data.
\protect\label{fig-xmaxvse}
}
\end{minipage}
\end{figure}

Calculations of average $X_{\rm max}$ values for different primaries
are shown in Figure~\ref{fig-xmaxvse}.
The large ``elongation rate'' (slope $dX_{\rm max} /d\lg E$)
for photons 
leads to $X_{\rm max}$ values well above those of nuclear primaries
already at $10^{17}$$-$$10^{18}$~eV.
Changes in the elongation rate at $10^{19}$$-$$10^{20}$~eV
are due to the LPM and preshower effects.

\begin{figure}[t]
\centering
\begin{minipage}[c]{.55\textwidth}
\centerline{\psfig{file=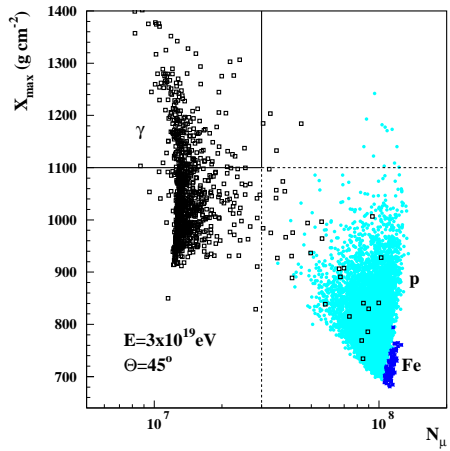,width=2.6in}}
\end{minipage}
\begin{minipage}[c]{.35\textwidth}
\centering
\caption{
$X_{\rm max}$ versus $N_{\mu}$ for different primaries
at $3 \times 10^{19}$~eV.
Simulations were performed for photons (1000 events),
protons (10000 events), and iron nuclei (100 events)
with PRESHOWER, CONEX and QGSJET 01.
No detector effects are included.
The lines illustrate possible cuts 
to achieve small misidentification rates
for photon identification (see text).
\protect\label{fig-xmaxnmu}}
\end{minipage}
\end{figure}

In addition to $X_{\rm max}$, showers from photon and nuclear primaries
differ in $N_{\mu}$ due to the small photonuclear cross-section.
To study the impact of shower fluctuations on photon identification,
a scatter plot of $X_{\rm max}$ versus $N_{\mu}$ is given
in Figure~\ref{fig-xmaxnmu} for different primaries simulated
at $3 \times 10^{19}$~eV.
This is an idealized plot as no detector effects are accounted for.
It can be seen that shower fluctuations by themselves do not
impose major limitations to identify photons.
Misidentification rates of $<$$10^{-3}$ ($<$$10^{-4}$)
using $X_{\rm max}$ ($N_{\mu}$)
seem possible for protons and are even smaller for nuclei
(numbers may depend on the hadron generator used for calculating
nuclear primaries).
In turn, an unconverted primary photon may hardly be distinguished
from a nuclear primary if in one of the first interactions,
a photonuclear interaction occurs (probability of order $\sim$1\%).
Contrary to $X_{\rm max}$, $N_{\mu}$ does not differ much between
converted and unconverted photons.\cite{ns_photons}
Further studies of photon showers can be found, for instance,
in Refs.\cite{coll1,coll2,coll3,coll4,coll5,coll6,coll7,coll8,bertou00}.

There are other observables, usually closely related to
$X_{\rm max}$ or $N_{\mu}$ or a combination of both, that 
could be used to distinguish photons from nuclear primaries.
These observables, some of which are specific for a certain detector type
and experimental configuration,
include the curvature of the shower front and the steepness
of the lateral distribution\cite{bertou00} of ground particles,
the risetime\cite{augerphoton} of the detector signal at a certain
core distance as well as the signal amplitude by itself.

\subsection{Detector response to photon-induced showers}
\label{sub-det}

The differences between showers from photons and nuclear primaries
can affect the detector acceptance to and the energy reconstruction
of UHE photons. As the effects depend on the detector type,
analysis approaches appropriate for the experimental
apparatus can be developed to minimize uncertainties from
the corrections.

Suppose a two-component composition of photons and protons with fluxes
$\Phi_\gamma (E)$ and $\Phi_p (E)$ at energy $E$. The ratio of photons
to protons, one measure of the photon fraction, is
then $f_{\gamma p} (E) = \Phi_\gamma(E) / \Phi_p(E)$.
Similarly, integral fluxes or fractions, additional primaries,
or the photon fraction to the total flux could be considered.

Regarding the detector acceptances $A_i(E)$ with $i=\gamma$ or $p$,
an acceptance ratio of $\epsilon (E) = A_\gamma(E) / A_p(E) < 1$
affects both the flux and fraction of registered photons.
Such a bias can occur for near-vertical photon showers
reaching ground before being fully developed.
For instance, the aperture of the HiRes-I telescope was found to be
$\sim$40\% reduced for photons at $E_{\rm GZK}$.\cite{desouza}
In a recent analysis of Auger hybrid data, a minimum zenith angle was
introduced to reduce the acceptance bias to photons.\cite{augerphoton}
Experimental conclusions about the photon {\it flux} require
knowledge of $A_\gamma(E)$ (usually well controlled for arrays using
appropriate selection cuts). For conclusions about the photon
{\it fraction}, knowledge of the relative acceptance $\epsilon (E)$
(possibly easier to calculate for fluorescence telescopes)
may be sufficient.
The latter approach has been applied in the analysis of Auger
hybrid data.\cite{augerphoton}

Reconstructing an energy $E_{\rm rec}$ from observables that, on
average, differ between photon and proton showers of same energy $E$
and assuming proton primaries only, leads to a misreconstruction of
photon energies.
Selecting events according to $E_{\rm rec}$ then introduces
a shift in energy scales between the primaries:
events with the same $E_{\rm rec}$
refer to protons of energy $E_{\rm rec} = E$,
but to photons of energy $E_{\rm rec} / g = E / g $
with $g \ne 1$.
Hence, the photon flux and fraction actually entering the
data sample are $\Phi_\gamma (E/g) \ne \Phi_\gamma (E)$ and 
$\Phi_\gamma (E/g) / \Phi_p (E) \ne f_{\gamma p} (E)$.
In general, a correction for this shift in energy scales
is required.

For the fluorescence technique using the integrated calorimetric
shower energy as an observable, $g \sim 1.1$, i.e. the energy scales
nearly match and $g>1$.
The small difference in energy scales is due to the missing energy correction,
which is larger for nuclear primaries compared to the almost purely
electromagnetic photon showers.\cite{pierog}
A conservative upper limit to the {\it integral} flux or fraction of photons
above $E$ may then be possible {\it without} correction or assumptions
on $\Phi_\gamma (E)$
since for $g\sim 1.1 \Rightarrow E/g<E
\Rightarrow \int_{E/g} \Phi_\gamma (E') dE' \ge \int_E \Phi_\gamma (E') dE'$.
This was used in Ref.~\cite{augerphoton}.

For arrays when reconstructing the energy from a signal
amplitude $S(r)$ measured at ground at core distance $r$,
the factor $g$ depends in general on the arrival direction of the
event (slant depth to ground changes with zenith, preshower
probability depends on zenith and azimuth), its energy (dependence
of $X_{\rm max}$ with energy, see Figure \ref{fig-xmaxvse}), and detector
characteristics (e.g. sensitivity to shower muons).
Values of $g \sim 0.5$ are possible.\cite{shinozaki,sommers}
Additionally, large shower fluctuations may occur
(for a proposal to reduce the effect of shower fluctuations
when reconstructing photon energies, see Ref.\cite{billoir07}).
The (for arrays larger) uncertainty when relating different
energy scales may be avoided by analysing the photon {\it flux}
rather than the fraction,
as for the flux, only the photon energy scale is required
(and $A_\gamma (E)$ is usually well controlled for arrays, see above).

Within the flux level experimentally allowed for photons, the different detector
response to photons introduces a systematic uncertainty when
reconstructing an energy spectrum.
An example of how the high-energy end of the spectrum from first
Auger data can change when simply assuming a constant factor two
between photon and proton energies is shown in Ref.\cite{busca}.
Detector responses to deep showers in general are discussed
in Ref.\cite{chou}.

\section{Experimental Searches for UHE Photons}
\label{sec-exp}

\subsection{Status}
\label{sec-status}

No photon detection has been reported so far.
Upper limits to photons come from different experiments,
see also Figure~\ref{fig-uplim}.

Comparing rates of near-vertical showers to inclined ones
recorded with the Haverah Park water detectors,
upper limits (95\% c.l.) of
48\% above $10^{19}$~eV (52 events) and 50\% above $4 \times 10^{19}$~eV
(10 events) were deduced.\cite{ave1,ave2}
From simulations, ground signals were found to be stronger suppressed
for photons than for nuclear primaries at large zenith angles. 
The absence of this suppression in the data gives constrains on the
photon contribution.
Contrary to other approaches, a feature of the data sample as a whole
is analysed here instead of observables in single events.

With muon counters in the AGASA array, the muon density $\rho_\mu$
at 1000~m core distance was measured for part of the events.\cite{shinozaki}
A simulation study of the ground signal S(600) for different
primaries showed a possible underestimation of photon energies
(e.g. about 30\%, 50\%, 20\% at
$10^{19}$~eV, $10^{19.5}$~eV, $>$$10^{20}$~eV).
Assuming a mixture of primary photons and protons,
limits (95\% c.l.) to the photon
fraction were estimated to be 28\% above $10^{19}$~eV (102 events)
and 67\% above $3.2 \times 10^{19}$~eV (14 events).\cite{shinozaki}
The data were compared to an overall simulated distribution,
i.e.~not event-by-event.
There seem to be some ``photon-like'' events
(not commented on in the paper):
in about five events, the measured $\rho_\mu$ is 
considerably (factor $\sim$4 or more)
below typical values expected for nuclear primaries; in
additional four events, the (at least two) muon detectors registered
no signal;
one event of $E_{\rm rec} \sim 7.5 \times 10^{19}$~eV (energy 
scale for nuclear primaries) was observed with a value $\rho_\mu$
about a factor $\sim$2.5 below an average expected from a data fit.

A comparison to individual events is most interesting at
highest energy where every second event could be a photon
(cf. Figure~\ref{fig-uplim}).
Observed event features can be compared to high-statistics
photon simulations, and the chance probability of photons
to generate such an event can be determined.
This way, for the $3.2 \times 10^{20}$~eV Fly's Eye event\cite{bird}
(see Figure~\ref{fig-profiles})
with $X_{\rm max}^{\rm obs} \sim 815 \pm 60$~g~cm$^{-2}$,
photon shower profiles (with $X_{\rm max}^{\gamma} \sim 937 \pm 26$~g~cm$^{-2}$)
were found to differ by $\sim$ 1.5$\sigma$ 
from the observed profile.\cite{halzen_fe,fe04}
While a photon origin can not be excluded, profiles
from nuclear primaries fit the data better.\cite{stanev95,fe04}

The method of comparing individual events was also applied
in Ref.\cite{risse05} to the
six highest energy AGASA events with observed $\rho_\mu$.
The energy scale for photons and the uncertainties of reconstructed
energy and $\rho_\mu$ were taken according to AGASA findings 
in Ref.\cite{shinozaki}.
For all six events, the muon densities from photons are (factors 2$-$7) below
the observed ones.
Photon predictions would be even more discrepant to data if smaller
energies (see below) were assumed.
A statistical approach for deriving limits from a small number
of events was developed which accounts for shower features changing
for each event and for non-Gaussian shower fluctuations.
An upper limit of 67\% (95\% c.l.) above $1.25 \times 10^{20}$~eV was derived.
Though not ruling out non-acceleration models, these scenarios
appear to have problems to consistently explain the AGASA data (no flux
suppression and no photon dominance).

A method for comparing measured and simulated shower observables such as
S(600) instead of using the reconstructed energy was given in 
Refs.\cite{troitsky1,troitsky2}.
For individual events, the corresponding photon energies are found by
requiring the simulated S(600) to fit the observed ones and
properly weighting the events.
This way, shower and detector fluctuations can more directly
be accounted for.
Detailed information about the detector is needed.\cite{troitsky1,troitsky2}

Applied to AGASA data, using response functions for S(600) and
information on the muon detector response from Ref.\cite{shinozaki},
photon energies of the six highest-energy events
were found in Ref.\cite{troitsky1} to be
(up to factor $\sim$2) below those expected from Ref.\cite{shinozaki}.
There appear to be differences between the findings of Refs.\cite{troitsky1}
and~\cite{shinozaki} in the expected S(600) for fixed energy which
may be connected to the use of different simulation codes.
Combining six AGASA\cite{shinozaki} and four
Yakutsk\cite{egorova} events with $E_{\rm rec} > 8 \times 10^{19}$~eV
(energy scale for nuclear primaries), a 36\% limit above $10^{20}$~eV
was derived (95\% c.l.).\cite{troitsky1}
There may be a sensitivity to the choice of $E_{\rm rec}$.
Reducing $E_{\rm rec}$, a more ``photon-like'' AGASA event
(see above) enters the sample.
For a discussion about combining data from these two experiments
which appear to have systematic differences in the reconstructed
flux spectra, see e.g. Ref.\cite{aloisio}.

Using the same method of Ref.\cite{troitsky1,troitsky2} and investigating
Yakutsk data only, preliminary (not yet published) limits of
12\% above $2 \times 10^{19}$~eV (50 events) and 22\% above 
$4 \times 10^{19}$~eV (95\% c.l.) were obtained.\cite{yakutsk}
In two events no muon signal was registered, another event
was classified as muon-poor.

All these limits refer to the {\it fraction} of photons
and come from ground arrays.
The fluorescence technique provides important cross-checks
due to smaller uncertainties from different energy scales
of the primaries (Section \ref{sub-det}).
Using $X_{\rm max}$ from
the direct observation of the shower profile with fluorescence
telescopes in hybrid events (i.e. registered by ground and fluorescence
detectors)
a limit of 16\% (95\% c.l.) above $10^{19}$~eV (29 events) was
obtained from first data taken at the Auger Observatory.\cite{augerphoton}
An approach was developed such that the limit does not rely on
assumptions about input spectra or composition.
A $\sim$2$\times$$10^{20}$~eV event registered with 
$X_{\rm max}^{\rm obs} \sim 821 \pm 36$~g~cm$^{-2}$
differs by $\sim$3$\sigma$ from the photon hypothesis
($X_{\rm max}^{\gamma} \sim 948 \pm 27$~g~cm$^{-2}$).

Steps were taken to analyse the HiRes stereo data set of about 50 events
above $4 \times 10^{19}$~eV.\cite{hiresphoton} No results
were reported so far.

In conclusion, current data from ground and fluorescence detectors
do not indicate a large (above $10^{19}$~eV)
or dominant (above $10^{20}$~eV)
photon contribution. Non-acceleration models are constrained
by existing limits though not ruled out as an explanation
of UHE cosmic rays (see also discussion in Ref.\cite{aloisio}).

\subsection{Prospects}
\label{sec-prosp}

\begin{figure}[t]
\centerline{\psfig{file=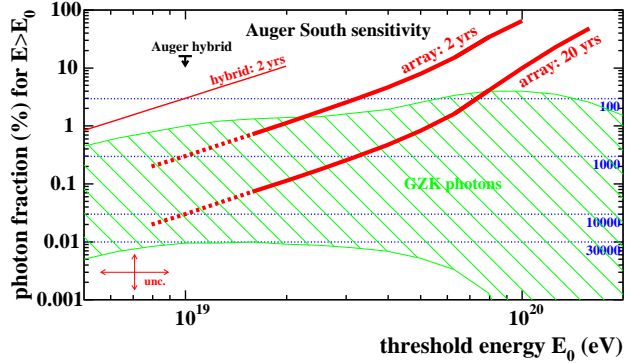,width=3.3in}}
\vspace*{8pt}
\caption{
Fraction of photons in the integral cosmic-ray flux as a
function of the threshold energy (see also Fig.~\ref{fig-uplim}).
The predictions\protect\cite{models} (labeled ``GZK photons'')
for primary nucleon sources follow a
spectrum with flux suppression above $E_{\rm GZK}$.
For reference, the current limit from Auger hybrid data,
obtained with 29 events, is shown.
The estimated sensitivity (see text) of the array
of the southern Auger Observatory
is shown for two and twenty years of operation for
a spectrum with flux suppression.
Also shown is the estimated sensitivity using hybrid data,
which are less numerous due to the $\sim$10\% duty cycle
but reach to smaller energy.
The uncertainty of the sensitivity estimates is
indicated at the lower left corner.
At around $10^{19}$~eV (dashed lines),
additional threshold effects may become increasingly important
for the array.
If complemented by an extended northern array, a sensitivity
level of below 0.1\% can be reached within a few years of
full operation.
\protect\label{fig-sens}}
\end{figure}

The 16\% limit from the Auger hybrid data is based on 29 events.
The sensitivity can be improved by a factor $\sim$3 or more
with data accumulated until 2008/2009.\cite{augerphoton}
Also energies below $10^{19}$~eV can be reached with the
hybrid technique.

For an overall estimate of future sensitivities
it is instructive
to consider the lowest theoretical fraction $F_\gamma^{\rm min}$ that
can be excluded with $n$ events. With $\alpha$
being the confidence level of rejection, $F_\gamma^{\rm min}$ is
\begin{equation}
\label{eq-sens}
F_\gamma^{\rm min}(n)  = 1-(1-\alpha)^{1/n}
\approx 3/n ~~
{\rm for} ~ \alpha = 0.95,~ n \gg 1 ~.
\end{equation}

A limit derived from data can exceed $F_\gamma^{\rm min}$ (i.e. be
weaker)
depending on the discrimination power of the specific observables
(possibly limited by shower fluctuations and detector characteristics),
efficiency corrections,
and the actual number of photons in the data sample.
For the example of the Auger hybrid limit based on 29 events,
$F_\gamma^{\rm min}(29) \sim 10\%$. The final limit of 16\%,
which includes a $\sim 20\%$ efficiency correction,
corresponds to the theoretical limit for 17 events
($F_\gamma^{\rm min}(17) \sim 16\%$).
Thus, compared to the 29 events in the data set, the 16\% limit
corresponds to an effective number of events that is smaller by a
factor $r \sim 29/17 \sim 1.7$.

To estimate the sensitivity to photons of the large Auger array,
we assume a factor of $r$$\sim$2.4, i.e. larger (less discriminative
events and/or larger efficiency correction) than the factor
reached in the first Auger hybrid limit.
Corresponding numbers from other event-by-event photon searches
using AGASA and Yakutsk data
are $r$$\sim$2.2,\cite{risse05} $\sim$2.1,\cite{yakutsk}
and $\sim$1.15 \cite{troitsky1}.
More detailed studies, particularly of threshold
effects, are needed. However, the uncertainty of this
estimate (factor $\sim$2) seems well below that of current theoretical
predictions of UHE photon fluxes.
Using for $n(E>E_0)$ the 
Auger energy spectrum\cite{sommers} gives the
sensitivity estimates $F(E>E_0)$
shown in Figures~\ref{fig-uplim} and \ref{fig-sens}
by calculating $F=F_\gamma^{\rm min}(n') \sim 3r/n$ 
where $n' = n/r$ denotes the effective number of 
events.\footnote{ With the integral number
of events $n(E>E_0)$, from Eq.~(\ref{eq-sens}) it follows
$ F_\gamma^{\rm min} \propto E_0 ^{\beta -1}$
for a power law spectrum of differential index $\beta$
($\beta \sim 2.84$ from Ref.~\cite{sommers}),
i.e.~a straight line of slope $\beta -1$ in
double-logarithmic scales. Zenith angles were restricted to
30$-$60$^\circ$ for the estimate of the sensitivity.}
The curve $F(E_0)$ indicates the upper limit that can be set
at 95\% c.l. if there were no photons in the data.
The curve represents also
(since $F(E_0) \sim 3r/n$) the ratio of 3$r \simeq 7$ to $n$ events:
if it is possible to identify 7 photons out of $n(E_0)$ events,
$F(E_0)$ gives the {\it observed} photon fraction.

A crucial test of non-acceleration models is possible with
the southern Auger array even if 
the flux is suppressed above $E_{\rm GZK}$ (Figure~\ref{fig-uplim}).
A large exposure is
required to reach sensitivities of $\sim 0.1\%$ or below, i.e. to
possibly observe GZK photons expected from UHE nucleon sources
(Figure~\ref{fig-sens}).
Such sensitivity levels may be reached on acceptable time scales
if the southern Auger Observatory is complemented by an expanded
(e.g. factor 3$-$4 larger) northern site.

The Telescope Array\cite{telarray} of scintillators, under
construction in Utah, will cover a $\sim$760~km$^2$ area
($\sim$25\% of the southern Auger array) overviewed by fluorescence
telescopes.
Space missions such as the proposed EUSO\cite{euso}
and OWL\cite{owl} projects may offer a significant increase of the 
experimental exposure to cosmic rays at highest energy.

\section{Possible Impact of UHE Photon Searches}
\label{sec-impact}

Photons, as the gauge bosons of the electromagnetic force,
at such enormous energy can be regarded as unique messengers and probes
of extreme and, possibly, new physics.
Implications are related to the production of photons, their propagation,
and interactions at the Earth.
Many aspects of the following, incomplete list
of possible impacts of UHE photon searches and connections
to other research subjects require more study.

Large UHE photon fluxes are a {\it smoking gun} for current
non-acceleration models.
Stringent photon limits give parameter constraints such as a lower
limit on the lifetime of relic SHDM particles.\cite{aloisio}

Findings on photons are needed to reduce corresponding systematics
in other air shower studies, such as for the energy spectrum
(Section \ref{sub-det}) or when trying to constrain interaction
parameters such as the proton-air cross-section\cite{belov,ulrich} from
showers.

UHE photons may be helpful for diagnostics of sources accelerating
nuclear primaries, as the photon fluxes from UHE hadron interactions
are expected to be connected with source features
such as type of primary, injection spectrum,
possible beam dump at the source, or source distribution
(see also Ref.\cite{models}).

UHE photons point back to the location of their production.
Possibly, the arrival directions of photons may correlate better
with the source direction than those of charged primaries.
There may be an enhanced UHE photon flux from the galactic center
region depending on the spectra of nuclear primaries.\cite{stecker06}
In certain SHDM scenarios, an enhanced flux of $\sim$$10^{18}$~eV
photons from the galactic center is possible without a higher-energy
counterpart.\cite{berez_priv}

Propagation features of UHE photons are sensitive to the
MHz radio background.\cite{sarkar03}
The photon flux at Earth is also
sensitive to extragalactic magnetic fields.\cite{sigl95}

Already a small sample of photon-induced showers may provide
relatively clean probes of aspects of QED and QCD at ultra-high energy
via the preshower process and photonuclear interactions
(Section \ref{sub-process}).

Certain implications will be most powerful in combination
with other results from shower observation (spectrum, anisotropy,
nuclear composition) and, probably, with results using other
messenger particles (UHE neutrinos, lower energy
photons).
Parallel to UHE photons, UHE neutrinos are usually 
produced, see Eq.~(\ref{eq-production}).
Due to the much longer mean free path of UHE neutrinos,
searches for UHE photons and UHE neutrinos complement each other,
with UHE photons (neutrinos) testing more local (distant)
production sites.
The disappearance of UHE photons during propagation
is accompanied
by an appearance of GeV-TeV photons, cf. Eq.~(\ref{eq-propagation})
and Figure \ref{fig-mfp} (see also Ref.\cite{armengaud} and
references therein).
Similar to (but independent of) UHE neutrinos,
these GeV-TeV photons allow a test of
more distant production sites.  The close relation between
different messengers is reflected by the fact
that constraints on non-acceleration models also come from
UHE neutrino and GeV photon data, see e.g.~Refs.\cite{berez04,sem-sigl}.
The full information could be exploited by
{\it multi-messenger observations}.
For instance, for known sources of UHE cosmic rays, fluxes
of UHE photons much below the expected level could indicate certain
new physics.

There are several connections to Lorentz invariance
violation.\cite{liv1,liv2}
The production of GZK photons can be affected
as well as interactions of photons during propagation and
when initiating a cascade at the Earth.
Particularly, photon conversion (interaction with
background fields or preshower process)
may be suppressed.

It is interesting to check whether UHE photon propagation
could be affected by the presence of axions or scalar bosons.
Formal requirements for photon conversion regarding photon
energy and magnetic field strength, appear to be 
fulfilled.\cite{gabrielli1,gabrielli2}
Photon conversions to non-electromagnetic channels
may differ from the standard QED process of Eq.~(\ref{eq-propagation})
due to an absence of an electromagnetic sub-cascade.

UHE photon propagation can be modified in certain models of
brane worlds,\cite{branes} quantum gravity theory,\cite{quantum1,quantum2}
or spacetime foam\cite{klinkhamer1,klinkhamer2}.
For instance, depending on fundamental length scales significant scattering
of photons on structures (defects) in spacetime foam can occur.
In turn, constraints may be derived when actually observing UHE photons,
even with one gold-plated event only.\cite{klinkhamer1,klinkhamer2}

\section{Conclusions}
\label{sec-summ}

UHE photons are tracers of highest-energy processes and new physics.
Showers initiated by such photons can well be distinguished from
those by nuclear primaries.
When complemented by a large northern site, the Pierre Auger Observatory
is expected to be sensitive to photon fractions of 0.1\% and below
and can realistically aim at photon observations.
The search for UHE photons contributes to multi-messenger observations
of the universe.
A key characteristic for the progress in astrophysics is
to expand to photon wavelengths beyond the optical;
the observation of UHE photons would bring this
to the highest end.

\section*{Acknowledgments}

We thank D. Barnhill, J. Bellido,
V. Berezinsky, P. Billoir, B. Dawson, K. Eitel,
R.~Engel, E. Gabrielli, D.~G\'ora, D. Gorbunov, J.-C. Hamilton,
M. Healy, D.~Heck,
M.~Kachelrie{\ss}, K.-H. Kampert, F.R. Klinkhamer, S. Ostapchenko,
J.~P\c{e}kala, M.~Roth, C. Roucelle,
S. Sarkar, D.~Semikoz, K. Shinozaki, G. Sigl, P. Sommers, M. Strikman,
S. Troitsky, M. Unger, S. Vorobiov, A.A. Watson,
B.~Wilczy\'nska, H.~Wilczy\'nski,
and all members 
of the Auger ``photon group'' for many useful discussions.
The simulations and plot routines for Figure~\ref{fig-xmaxvse}
were kindly provided by D. Heck.
This work was partially supported by the Polish Ministry of Science
and Higher Education (Grant N202 090 31/0623)
and by the German Ministry for Research and Education
(Grant 05 CU5PX1/6).
One of the authors (MR) kindly acknowledges support from the
Alexander von Humboldt foundation.
{\it Preprint of a brief review submitted for consideration in
Modern Physics Letters A (copyright World Scientific Publishing
Company, 2007, http://www.worldscientific.com.sg).}



\begin{thebibliography}{0}

\bibitem{weekes}
See, for instance, T.C. Weekes, [arXiv:astro-ph/0606130], and references therein.

\bibitem{auger}
Pierre Auger Collaboration, Nucl. Instr. Meth. {\bf A 523}, 50 (2004).

\bibitem{reviews_cr1}
M.~Nagano, A.A.~Watson, Rev.~Mod.~Phys. {\bf 72}, 689 (2000).
\bibitem{reviews_cr2}
L. Anchordoqui {\it et al.}, Int. J. Mod. Phys. A {\bf 18}, 2229 (2003).
\bibitem{reviews_cr3}
{\it ``Ultimate energy particles in the Universe''}, eds.
M.~Boratav and G.~Sigl, C.R.~Physique {\bf 5}, Elsevier, Paris (2004).
\bibitem{reviews_cr4}
J.~Cronin, Nucl.~Phys.~B, Proc.~Suppl. {\bf 138}, 465 (2005).

\bibitem{reviews_nu1}
F.~Halzen, D. Hooper, Rep. Prog. Phys. {\bf 65}, 1025 (2002).
\bibitem{reviews_nu2}
A.B. McDonald {\it et al.}, Rev. Sci. Instrum. {\bf 75}, 293 (2004).

\bibitem{bhat-sigl}
P.~Bhattacharjee, G.~Sigl, Phys.~Rep.~{\bf 327}, 109 (2000).

\bibitem{td1}
C.T.~Hill, Nucl.~Phys.~B {\bf 224}, 469 (1983).
\bibitem{td2}
M.B.~Hindmarsh, T.W.B.~Kibble, Rep.~Prog.~Phys.~{\bf 58}, 477 (1995).

\bibitem{shdm1}
V.~Berezinsky, M.~Kachelrie{\ss}, A.~Vilenkin,
Phys.~Rev.~Lett.~{\bf 79}, 4302 (1997).
\bibitem{shdm2}
M.~Birkel, S.~Sarkar, Astropart.~Phys.~{\bf 9}, 297 (1998).
\bibitem{shdm3}
V.A. Kuzmin, V.A. Rubakov, Phys. Atom. Nucl. {\bf 61}, 1028 (1998).
\bibitem{shdm4}
P. Blasi, R. Dick, E.W. Kolb, Astropart.~Phys.~{\bf 18}, 57 (2002).

\bibitem{frag1}
Z.~Fodor, S.D.~Katz, Phys.~Rev.~Lett.~{\bf 86}, 3224 (2001).
\bibitem{frag2}
S.~Sarkar, R.~Toldra, Nucl.~Phys.~B {\bf 621}, 495 (2002).
\bibitem{frag3}
C.~Barbot, M.~Drees, Astropart.~Phys.~{\bf 20}, 5 (2003).

\bibitem{berez04}
R.~Aloisio, V.~Berezinsky, M.~Kachelrie{\ss}, Phys. Rev. D {\bf 69}, 094023 (2004).

\bibitem{sarkar03}
S.~Sarkar, Acta Phys.~Polon.~{\bf B35}, 351 (2004).

\bibitem{models}
G.~Gelmini, O.E.~Kalashev, D.V.~Semikoz, [arXiv:astro-ph/0506128].

\bibitem{ellis}
J.~Ellis, V.~Mayes, D.V.~Nanopoulos, Phys. Rev. D {\bf 74}, 115003 (2006).

\bibitem{zb1}
T.J.~Weiler, Phys.~Rev.~Lett.~{\bf 49}, 234 (1982).
\bibitem{zb2}
T.J.~Weiler, Astropart.~Phys.~{\bf 11}, 303 (1999).
\bibitem{zb3}
D.~Fargion, B.~Mele, A.~Salis, Astrophys.~J.~{\bf 517}, 725 (1999).

\bibitem{gzk1}
K.~Greisen, Phys.~Rev.~Lett.~{\bf 16}, 748 (1966).
\bibitem{gzk2}
G.T.~Zatsepin, V.A.~Kuzmin, JETP Lett.~{\bf 4}, 78 (1966).

\bibitem{stecker06}
A. Kusenko, J. Schissel, F.W. Stecker,
Astropart. Phys. {\bf 25}, 242 (2006).

\bibitem{losslengths1}
S. Lee, Phys. Rev. D {\bf 58}, 043004 (1998).
\bibitem{losslengths2}
D. Hooper, S. Sarkar, A.M. Taylor, 
Astropart. Phys. 27 (2007) 199.

\bibitem{hires-gzk}
R.U.~Abbasi {\it et al.}, Phys.~Lett.~B {\bf 619}, 271 (2005).

\bibitem{agasa-gzk}
M.~Takeda {\it et al.}, Astropart.~Phys. {\bf 19}, 447 (2003).

\bibitem{aloisio}
R. Aloisio, V. Berezinsky, M.~Kachelrie{\ss},
Phys.~Rev.~D {\bf 74}, 023516 (2006).

\bibitem{augerphoton}
Pierre Auger Collaboration, Astropart. Phys. {\bf 27}, 155 (2007).

\bibitem{ave1} 
M.~Ave {\it et al.}, Phys.~Rev.~Lett.~{\bf 85}, 2244 (2000).
\bibitem{ave2}
Phys.~Rev.~D {\bf 65}, 063007 (2002).

\bibitem{shinozaki} 
K.~Shinozaki {\it et al.}, Astrophys.~J.~{\bf 571}, L117 (2002).

\bibitem{risse05} 
M.~Risse {\it et al.}, Phys.~Rev.~Lett. {\bf 95}, 171102 (2005).

\bibitem{yakutsk}
A.V. Glushkov {\it et al.}, [arXiv:astro-ph/0701245v1].

\bibitem{troitsky1}
G.I.~Rubtsov {\it et al.}, Phys.~Rev.~D~{\bf 73}, 063009 (2006).
\bibitem{troitsky2}
D.S. Gorbunov, G.I. Rubtsov, S.V. Troitsky, [arXiv:astro-ph/0606442].

\bibitem{sigl06}
G. Sigl, [arXiv:astro-ph/0612240].

\bibitem{presh1}
T.~Erber, Rev.~Mod.~Phys.~{\bf 38}, 626 (1966)
(see Refs. \cite{cpc,klein1,klein2} for a caveat related to Table VI of the paper);
\bibitem{presh2}
V.H. Bayer, B.M. Katkov, V.S. Fadin, Radiation of Relativistic Electrons (in Russian),
Atomizdat, Moscow (1973);
\bibitem{presh3}
V.B.~Berestetskii {\it et al.},
Quantum Electrodynamics, Pergamon Press, 2nd edition (1982);
\bibitem{presh4}
A.A. Sokolov, I.M. Ternov, Radiation from Relativistic Electrons,
Springer Verlag (1986);
\bibitem{presh5}
M.G. Baring, A\&A {\bf 225}, 260 (1989);
\bibitem{presh6}
U.I. Uggerh\o j, Nucl. Phys. B (Proc. Suppl.) {\bf 122}, 357 (2003).

\bibitem{daugherty83}
J. K. Daugherty, A. K. Harding, Astrophys. J. {\bf 273}, 761 (1983).

\bibitem{ns_photons}
P.~Homola {\it et al.}, Astropart. Phys. 27 (2007) 174.

\bibitem{coraddu02}
M. Coraddu {\it et al.}, [arXiv:hep-ph/0210107].

\bibitem{cpc}
P.~Homola {\it et al.}, Comput.~Phys.~Commun. {\bf 173}, 71 (2005).

\bibitem{heck}
D.~Heck {\it et al.}, Reports {\bf FZKA 6019 \& 6097},
Forschungszentrum Karls\-ruhe (1998).

\bibitem{conex}
T.~Bergmann {\it et al.}, Astropart. Phys. {\bf 26}, 420 (2006).

\bibitem{magics}
D. Badagnani, S.J. Sciutto, Proc. 29$^{\rm th}$ Intern.~Cosmic Ray Conf.,
Pune, {\bf 9}, 1 (2005).

\bibitem{aires}
S.J. Sciutto, [arXiv:astro-ph/9911331].

\bibitem{igrf}
National Geophysical Data Center, USA, http://www.ngdc.noaa.gov.

\bibitem{bh}
H.A.~Bethe, W.~Heitler, Proc.~Roy.~Soc., {\bf A146} 83 (1934).

\bibitem{pdg06}
W.M. Yao {\it et al.}, J. Phys. G: Nucl. Part. Phys. {\bf 33}, 1 (2006).

\bibitem{lanpom53a} 
L.D.~Landau, I.Ya.~Pomeranchuk,
Dokl. Akad. Nauk SSSR {\bf 92}, 535 (1953).
\bibitem{lanpom53b}
L.D.~Landau, I.Ya.~Pomeranchuk,
Dokl. Akad. Nauk SSSR {\bf 92}, 735 (1953).

\bibitem{mig56}
A.B.~Migdal, Phys. Rev. {\bf 103}, 1811 (1956).

\bibitem{klein1}
S.~Klein, Rev.~Mod.~Phys. {\bf 71}, 1501 (1999).
\bibitem{klein2}
S.~Klein, Rad.~Phys.~Chem. {\bf 75}, 696 (2006).

\bibitem{hansen}
H.D. Hansen {\it et al.}, Phys. Rev. Lett. {\bf 91}, 014801 (2003).

\bibitem{cillis}
A.N. Cillis {\it et al.}, Phys. Rev. D {\bf 59}, 113012 (1997).

\bibitem{sib21}
R.~Engel {\it et al.},
Proc.~26$^{\rm th}$ Intern.~Cosmic Ray Conf., Salt Lake City, 415 (1999).

\bibitem{qgs01}
N.N. Kalmykov, S.S. Ostapchenko, A.I. Pavlov,
Nucl. Phys. B (Proc. Suppl.) {\bf 52}, 17 (1997).

\bibitem{qgs2}
S.~Ostapchenko,
Nucl. Phys. B (Proc. Suppl.) {\bf 151}, 143 (2006).

\bibitem{rissec2cr}
M.~Risse {\it et al.}, Czech.~J.~Phys. {\bf 56}, A327 (2006).

\bibitem{strikman}
T.C.~Rogers, M.I.~Strikman, J. Phys. G: Nucl. Part. Phys.
{\bf 32}, 2041 (2006).

\bibitem{pdg1}
S.~Eidelmann {\it et al.}, Phys.~Lett.~B {\bf 592}, 1 (2004).
\bibitem{pdg2}
J.R.~Cudell {\it et al.}, Phys.~Rev.~D {\bf 65}, 074024 (2002).

\bibitem{donland}
A.~Donnachie, P.~Landshoff, Phys.~Lett.~B {\bf 518}, 63 (2001).

\bibitem{bird}
D.J.~Bird {\it et al.}, Astrophys.~J.~{\bf 441}, 144 (1995).

\bibitem{fe04}
M.~Risse {\it et al.}, Astropart.~Phys.~{\bf 21}, 479 (2004).

\bibitem{halzen_fe}
F.~Halzen, [arXiv:astro-ph/0302489].

\bibitem{xmax-heck} J.~Knapp {\it et al.},
Astropart.~Phys.~{\bf 19}, 77 (2003).

\bibitem{coll1}
B.~McBreen, C.J.~Lambert, Phys.~Rev.~D {\bf 24}, 2536 (1981).
\bibitem{coll2}
F.A.~Aharonian, B.L.~Kanevsky, V.V.~Vardanian,
Astrophys.~Space Sci.~{\bf 167}, 111 (1990).
\bibitem{coll3}
H.P.~Vankov, P.V.~Stavrev, Phys.~Lett.~B {\bf 266}, 178 (1991).
\bibitem{coll4}
T.~Stanev, H.P.~Vankov, Phys.~Rev.~D {\bf 55}, 1365 (1997).
\bibitem{coll5}
A.V.~Plyasheshnikov, F.A.~Aharonian, J.~Phys.~G {\bf 238}, 267 (2002).
\bibitem{coll6}
J.N. Capdevielle, C. Le Gall, Kh.N. Sanosyan,
Astropart. Phys. {\bf 13}, 259, (2000).
\bibitem{coll7}
W.~Bednarek, New Astronomy {\bf 7}, 471 (2002).
\bibitem{coll8}
H.P.~Vankov, N.~Inoue, K.~Shinozaki, Phys.~Rev.~D {\bf 67},
043002 (2003).

\bibitem{bertou00}
X.~Bertou, P.~Billoir, S.~Dagoret-Campagne,
Astropart.~Phys. {\bf 14}, 121 (2000).

\bibitem{desouza}
V.~de Souza, G.~Medina-Tanco, J.A. Ortiz, Phys.~Rev.~D~{\bf 72},
103009 (2005).

\bibitem{pierog}
T.~Pierog {\it et al.}, Proc. 29$^{\rm th}$ Intern.~Cosmic Ray Conf.,
Pune, {\bf 7}, 103 (2005).

\bibitem{sommers} P.~Sommers for the Auger Collaboration,
Proc. 29$^{\rm th}$ Intern.~Cosmic Ray Conf., Pune,
{\bf 7}, 387 (2005).

\bibitem{billoir07}
P. Billoir, C. Roucelle, J.-C. Hamilton, [arXiv:astro-ph/0701583].

\bibitem{busca}
N. Busca, D. Hooper, E.W. Kolb, Phys.~Rev.~D {\bf 73}, 123001 (2006).

\bibitem{chou}
A.S. Chou, Phys.~Rev.~D {\bf 74}, 103001 (2006).

\bibitem{stanev95}
F.~Halzen {\it et al.}, Astropart.~Phys. {\bf 3}, 151 (1995).

\bibitem{egorova}
V. Egorova {\it et al.}, Nucl. Phys. B (Proc. Suppl.)
{\bf 136}, 3 (2004).

\bibitem{hiresphoton}
HiRes Collaboration,
Proc. 29$^{\rm th}$ Intern.~Cosmic Ray Conf., Pune {\bf 7}, 373 (2005).

\bibitem{footnote} With the integral number
of events $n(E>E_0)$, from Eq.~(\ref{eq-sens}) it follows
$ F_\gamma^{\rm min} \propto E_0 ^{\beta -1}$
for a power law spectrum of differential index $\beta$,
($\beta \sim 2.84$ from Ref.~\cite{sommers}),
i.e.~a straight line of slope $\beta -1$ in
double-logarithmic scales. Zenith angles were restricted to
30$-$60$^\circ$ for the estimate of the sensitivity.

\bibitem{telarray}
M. Fukushima {\it et al.}, Prog.~Theor.~Phys.~Suppl.~{\bf 151}, 206
(2003).

\bibitem{euso}
http://www.euso-mission.org

\bibitem{owl}
http://owl.gsfc.nasa.gov

\bibitem{belov}
K. Belov et al., Nucl. Phys. B (Proc. Suppl.) {\bf 151}, 197 (2006).
\bibitem{ulrich}
R. Ulrich et al., [arXiv:astro-ph/0612205].

\bibitem{berez_priv}
V. Berezinsky, private communication (2005).

\bibitem{sigl95}
S. Lee, A.V. Olinto, G. Sigl, Astrophys. J. {\bf 455}, L21 (1995).

\bibitem{armengaud}
E. Armengaud, G. Sigl, F. Miniati, Phys. Rev. D {\bf 73}, 083008 (2006).

\bibitem{sem-sigl}
D.V.~Semikoz, G.~Sigl, JCAP {\bf 0404}, 003 (2004).

\bibitem{liv1}
T. Jacobson, S. Liberati, D. Mattingly, Annals Phys. {\bf 321}, 150 (2006).
\bibitem{liv2}
B. Altschul, [arXiv:hep-ph/0610324].

\bibitem{gabrielli1}
E. Gabrielli, K. Huitu, S. Roy, Phys. Rev. D {\bf 74}, 073002 (2006).
\bibitem{gabrielli2}
E. Gabrielli, private communication (2006).

\bibitem{branes}
M. Gogberashvili, A.S. Sakharov, E.K.G. Sarkisyan,
Phys. Lett. B {\bf 644}, 179 (2007).

\bibitem{quantum1}
G. Amelino-Camelia {\it et al.}, Nature {\bf 393}, 763 (1998).
\bibitem{quantum2}
R. Gambini, J. Pullin, Phys. Rev. D {\bf 59}, 124021 (1999).

\bibitem{klinkhamer1}
S. Bernadotte, F.R. Klinkhamer, Phys. Rev. D {\bf 75}, 024028 (2007).
\bibitem{klinkhamer2}
F.R. Klinkhamer, private communication (2007)

\end{thebibliography}
\end{document}